# Numerical Analysis of Damage Evolution in Open-Hole CFRP Laminates Modified with Electrospun Self-Healing Diels–Alder Interleaves

Marianna Chantzi[1], Vassilis Kostopoulos[1] and Spyridon Psarras[1]*

[1]Department of Mechanical Engineering and Aeronautics, University of Patras, GR-26500, Patras, Greece; corresponding author: spsarras@upatras.gr

**Abstract:**

This study presents a detailed finite element analysis of open-hole carbon fiber-reinforced polymer (CFRP) laminates modified with electrospun interleaves containing Diels–Alder-based self-healing agents. Building on prior experimental work, we develop a high-fidelity simulation framework to investigate the quasi-static tensile behavior of these advanced composites. Hashin's failure criteria are used to capture intralaminar damage, while surface-based cohesive contact interactions model interlaminar delamination. Two interleave configurations are examined: a solution electrospinning process (SEP) providing full-thickness coverage and a melt electrospinning process (MEP) offering localized reinforcement. Results show good agreement with experimental data, capturing key failure mechanisms such as matrix cracking, fiber breakage, and delamination. SEP-modified laminates exhibited enhanced toughness and load capacity, while MEP-modified specimens showed more localized, controlled damage. The study highlights the importance of spatially resolved cohesive properties and meshing strategies in accurately simulating damage progression. These findings provide valuable insights for designing and optimizing self-healing aerospace composites and form the basis for future fatigue and multifunctional performance simulations.

## 1 Introduction

In the past few decades, carbon fiber-reinforced polymers (CFRPs) have become a cornerstone in high-performance structural applications, particularly in the aerospace and automotive industries.

Their outstanding strength-to-weight ratio, tailorability, and corrosion resistance make them ideal for replacing conventional metallic components in weight-sensitive structures [1]. Typically manufactured as laminated composites with varying ply orientations, CFRPs exhibit pronounced anisotropy, which allows for optimization of mechanical performance along preferred directions. However, this same anisotropic nature contributes to their vulnerability to failure mechanisms such as matrix cracking, fiber breakage, and—most critically—interlaminar delamination [2].

Delamination, defined as the separation of adjacent plies within the laminate, is a pervasive issue in multilayered composites. It can originate from manufacturing defects, in-service impacts, or thermomechanical fatigue. When microcracks in the matrix coalesce under cyclic loading or stress concentrations, they give rise to delaminations that severely reduce the load-bearing capacity and structural integrity of the laminate [3]. Such failures are especially problematic in aerospace structures where safety and reliability are paramount.

One of the most commonly encountered structural features in composite components is the open hole (OH), which serves as an attachment point for fasteners or access points for maintenance. However, the presence of notches or holes introduces complex stress fields and amplifies local damage mechanisms, making the composite more susceptible to premature failure [4, 5] . The so-called notch sensitivity of CFRPs depends on several factors including laminate stacking sequence, notch size, ply thickness, and machining quality [6-8]. As such, the evaluation of open-hole tensile (OHT) performance remains a critical step in the qualification and design of composite airframe components [9-11].

To better understand and predict failure in OHT specimens, numerical modeling has emerged as a powerful and necessary complement to experimental testing. Among the available approaches, cohesive zone modeling (CZM) has become widely adopted due to its ability to simulate both initiation and growth of delamination without the need for predefined cracks. CZM characterizes the fracture process zone (FPZ) via traction-separation laws (TSLs), which relate interface stresses to displacement jumps across plies[12-15]. One significant advantage of CZM is its versatility in modeling different fracture modes (mode I, II, or mixed-mode) within a unified damage framework. In particular, surface-based cohesive contact formulations—readily implemented in commercial finite element tools like Abaqus—offer efficient and mesh-independent ways to capture delamination phenomena without explicit interface elements [16-19].

Despite these advancements, numerical predictions often fall short of capturing the full complexity of failure mechanisms observed experimentally. For instance, stress redistributions due to longitudinal matrix splitting, or the influence of microstructural heterogeneities, are often oversimplified. Recent studies suggest that accurate simulation of matrix cracking requires mesh alignment along fiber directions and, in some cases, explicit representation of splitting routes to mimic real crack paths [20, 21] .

Parallel to advances in modeling, significant effort has been devoted to improving the intrinsic damage tolerance of CFRPs. One promising route involves the use of self-healing agents (SHAs)—typically thermally reversible polymers based on the Diels–Alder (DA) reaction—which are introduced into the laminate during fabrication. These SHAs enable the material to recover part of its mechanical performance after damage, thereby extending its service life and reducing maintenance costs. When combined with electrospinning techniques and carbon-based nanofillers, such as graphene nanoplatelets (GNPs), these healing systems can improve both toughness and healing efficiency[22, 23].

Building on previous experimental work [24], the present study investigates the use of finite element modeling (FEM) to simulate quasi-static open-hole tensile behavior in CFRPs modified with electrospun SHAs. Two electrospinning methods—solution (SEP) and melt (MEP)—were used to introduce the healing agents in selected interlaminar regions. The study employs Hashin's failure criteria to model fiber and matrix damage, while inter-ply delaminations are captured using surface-based cohesive contact interactions. The objective is to assess the predictive capability of the FEM models by comparing them against experimental results and to gain further insight into damage progression and failure mechanisms.

## 2 Materials And Methods

Model Geometry and Setup

To simulate the quasi-static tensile behavior of open-hole CFRP laminates with ply layout $[+45/-45/0/90]_{2S}$, Figure 1(a), three-dimensional finite element models were developed using the Abaqus/Explicit solver, Figure 1 (b). Each specimen geometry was modeled to match the experimental configuration incorporating a central circular hole to replicate the open-hole condition. The boundary conditions and loading setup were applied to reflect the actual test scenario, including constrained displacements at one end and a prescribed displacement at the opposite end.

Material Properties and Failure Criteria Each unidirectional ply was modeled as a transversely isotropic material, using mechanical properties sourced from manufacturer datasheets (SIGRAPREG C U150-0/NF-E340/38%) [23] and supplemented with literature values[24]. The density and the elastic material properties used for the modeling of each ply are summarized in Table 1.

In Figure 1, the assembled laminate for the open-hole models is shown. The partitions on both ends of the laminate represent the parts which are constrained by the machine grips. The partitions at the center of the plate indicate the area where SHA was added by means of electrospinning. Apart from the laminate, a tab was also modeled as a discrete rigid part in order to simulate the moving grip of the machine. No material was defined for the tab. The tab is connected to the laminate by means of tie constraint. The displacement is applied to the middle point of the tab (Reference Point). To the left end of the laminate all movements and rotations have been constrained. The compliance of the testing machine was considered negligible.

*Table 1: Density and elastic material properties of lamina ply (CE 1007 150-38)*

| UD CFRP Properties (SIGRAPREG C U150-0/NF-E340/38%) | |
|---|---|
| Ply Thickness (mm) | 0.14 |
| Density $\rho$ ($g/cm^3$) | 1.55 |
| Modulus in fiber direction $E_1$ (GPa) | 105 |
| Transverse moduli $E_2 = E_3$ (GPa) | 8.4 |
| In-plane Shear moduli $G_{12} = G_{13}$ (GPa) | 5.2 |
| Transverse Shear modulus $G_{23}$ (GPa) | 3 |
| Major Poisson's ratios $v_{12} = v_{13}$ | 0.318 |
| Minor Poisson's ratio $v_{23}$ | 0.4 |

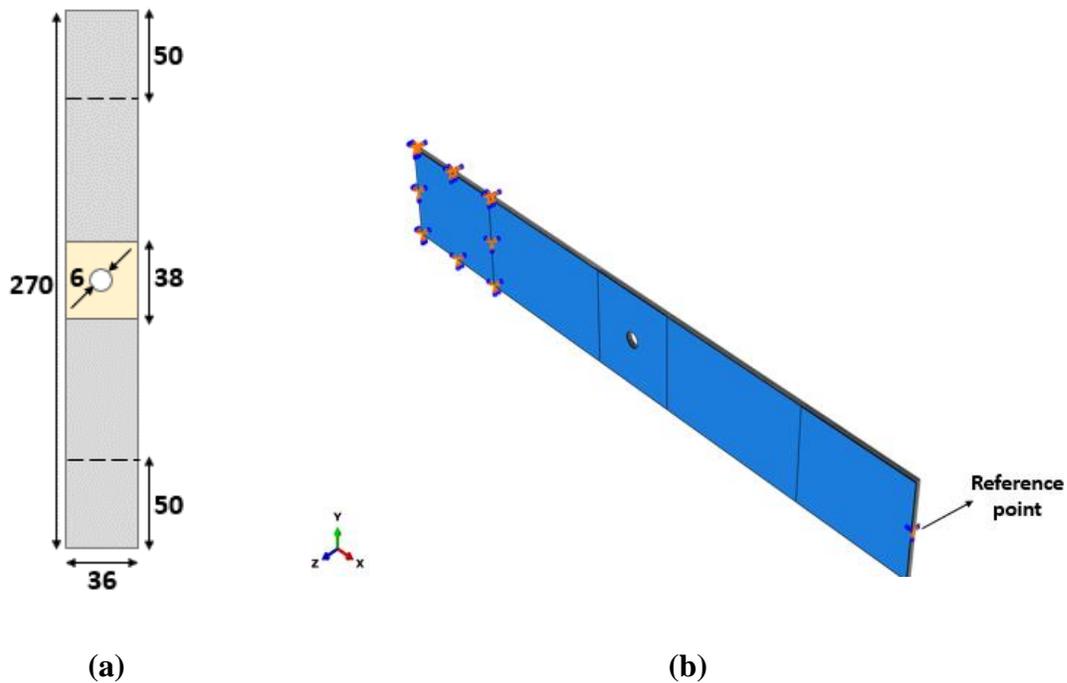

**(a)**          **(b)**

*Figure 1 (a) Configuration and dimensions of the unnotched (left) and open-hole (right) tension specimens. The lightly colored regions indicate where the SHA was added by means of electrospinning process. (b) Open-hole model with boundary conditions: encastre (left) at laminate partition and Displacement (right) at Tab Reference Point*

Hashin's failure criteria were employed to predict intralaminar damage, including fiber tension, fiber compression, matrix tension, and matrix compression. These criteria were combined with a stiffness degradation scheme to capture progressive damage. The use of a stiffness degradation model allowed for the gradual reduction of material properties as damage evolved, enabling a realistic simulation of failure modes observed in the experimental tests.

For interlaminar delamination, surface-based cohesive contact interactions were defined between plies. Separate cohesive property sets were assigned to modified and unmodified regions, corresponding to the presence or absence of the electrospun self-healing agent. These cohesive interactions were governed by a traction-separation law calibrated through preliminary mode I fracture tests [24], enabling the model to capture both initiation and propagation of delamination accurately.

Electrospinning Variants and Cohesive Modeling

Two self-healing configurations were studied: one using solution electrospinning (SEP), where the healing agent was introduced between all plies, and another using melt electrospinning (MEP), where the agent was selectively applied at [-45°] ply interfaces, Figure 2. In the numerical model, this distinction was implemented by assigning different cohesive properties to these zones. For SEP models (Figure 3) a uniform distribution of enhanced cohesive properties was applied throughout the laminate, while for MEP models (Figure 4) cohesive enhancement was localized. This approach captured the influence of SHA distribution patterns on delamination resistance and overall laminate performance.

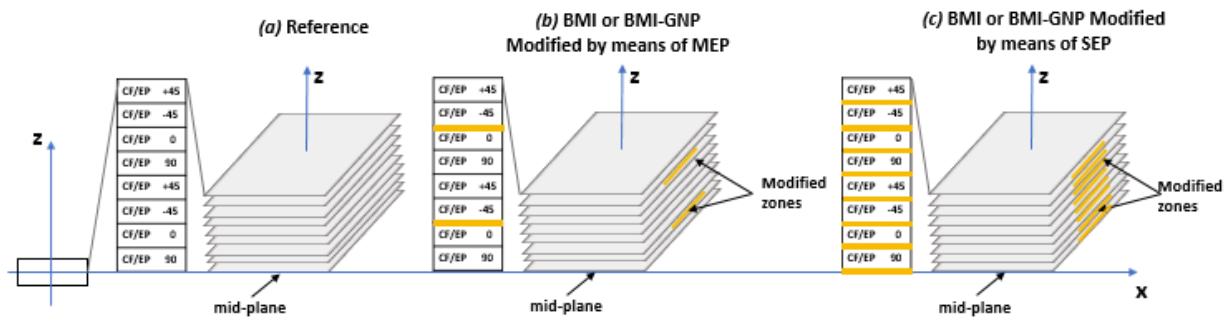

*Figure 2 Design of the (a) reference, (b) modified plates by means of MEP, (c) modified plates by means of SEP.*

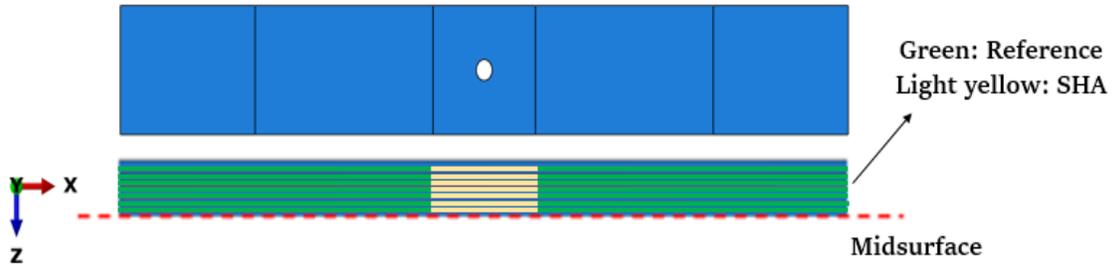

*Figure 3: Open-hole tensile model for SEP modified specimen: front view (top) and cohesive contact zones (bottom) highlighted in green for the reference material and light yellow for the SHA material. There are 30 cohesive surfaces in total (15 for the reference and 15 for the SHA cohesive interactions).*

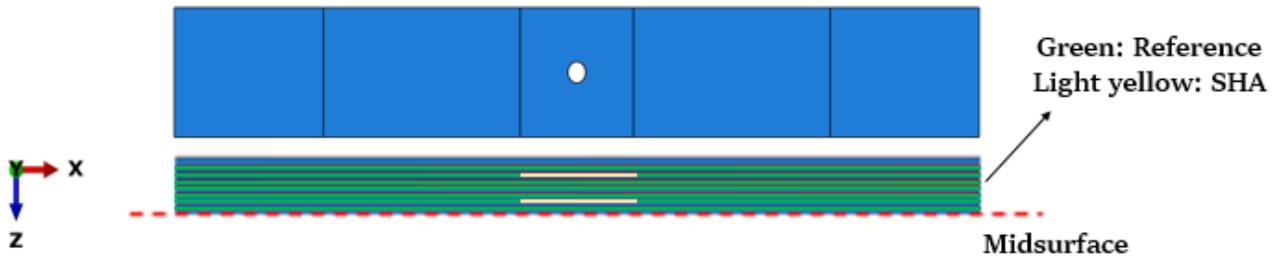

*Figure 4: Open-hole tensile model for MEP modified specimen: front view (top) and cohesive contact zones (bottom) highlighted in green for the reference material and light yellow for the SHA material. There are 19 cohesive surfaces in total (15 for the reference and 4 for the SHA cohesive interactions).*

Meshing Strategy and Convergence Study

Laminates were discretized using SC8R continuum shell elements, while rigid tabs were modeled with R3D4 elements to simulate gripping conditions, Figure 5. A mesh refinement study was conducted to identify an optimal mesh density that balances computational cost with predictive accuracy, Figure 6. Mesh sizes ranging from 2 mm to 4 mm were evaluated, Table 2. The 2 mm mesh exhibited the best correlation with experimental data in terms of peak load and damage evolution, particularly around the stress concentration region near the open hole. Further refinement was deemed unnecessary due to diminishing returns in result fidelity and a significant increase in computational time.

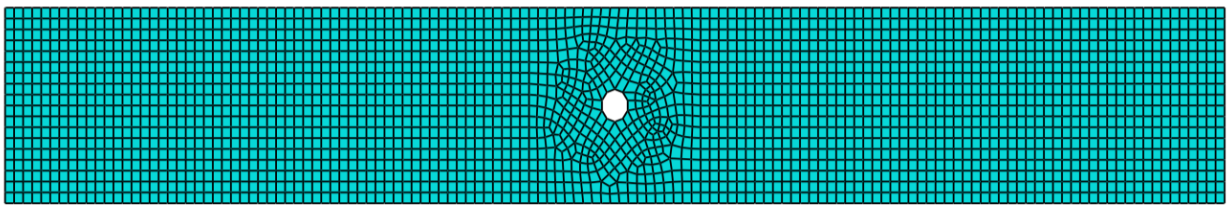

*Figure 5 Meshed OH specimen model*

Table 2 Open-hole tensile model: global mesh sizes and total analysis time for each investigated mesh

| Mesh label | Global mesh size | Analysis time |
|---|---|---|
| - | [mm] | [hrs] |
| M.2 | 2 | 5 |
| M.2,5 | 2.5 | ≈ 4 |
| M.3 | 3 | 4 |
| M.4 | 4 | 2 |

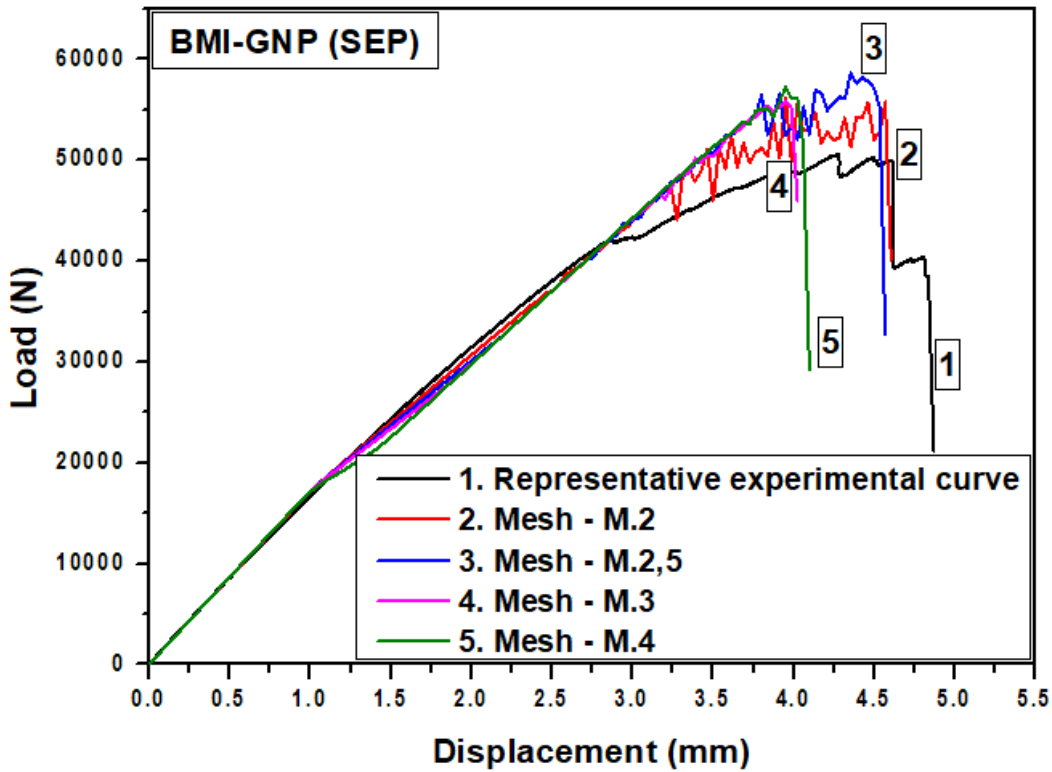

Figure 6 Mesh study: Load-displacement curves mesh size vs representative experimental curve.

Boundary Conditions and Loading

The models were subjected to quasi-static uniaxial tensile loading through a displacement-controlled boundary condition. One end of the laminate was fully constrained, while the opposite end was connected to a rigid tab subjected to a gradually increasing displacement. The loading rate was chosen to replicate experimental strain rates while maintaining numerical stability. The simulation maintained low kinetic energy relative to internal energy, ensuring quasi-static conditions and minimizing inertial effects.

<u>Calibration and Validation</u>

Cohesive and material damage parameters were fine-tuned through iterative comparison with experimental data. Fracture energies, damage initiation strengths, and stiffness properties were adjusted within experimentally justified ranges to achieve close correspondence between numerical and observed damage progression. Validation was performed through direct comparison of load-displacement curves and damage morphologies obtained from finite element simulations and those observed in actual test specimens. Specific attention was paid to the timing and location of delamination onset and matrix cracking, as these served as indicators of model fidelity.

The normal fracture energy used in the definition of the cohesive contact interaction was calculated in mode I experiments in [9] for the reference, BMI modified by means of SEP, and BMI-GNP modified by means of SEP material types. For the BMI and BMI-GNP modified by means of MEP materials the same fracture toughness values were used as in their SEP counterparts, Table 3

Regarding the properties needed for the definition of Hashin's criterion the datasheet provided only the longitudinal tensile strength and the interlaminar shear strength (2400 MPa and 90 MPa respectively). The rest of the properties were obtained through trial and error by comparing the analytical with the experimental results. Again, it was observed that the given longitudinal tensile strength (2400 MPa) gave a much lower peak load when compared to the experiments. Thus, after trials it was decided to set the value to 2750 MPa.

For the cohesive contact interaction, a friction and normal model were included in the cohesive contact property definition and "hard" contact and a friction coefficient of 0.3 were selected. The cohesive stiffnesses were chosen as the default values calculated by Abaqus. The final Hashin's criterion and cohesive contact interaction properties used for all material types are displayed in Table 4 to Overall, the modeling framework established in this study combines high-resolution finite element analysis with realistic material and interface behavior, enabling robust predictions of failure in self-healing open-hole CFRP laminates under tensile loading. The following sections present and analyze the simulation results in the context of the observed experimental phenomena.

Table 7. As discussed previously, two contact interaction properties were defined for the MEP modified models. In the tables REF refers to the zones with pristine material and SHA to the zones where SHA was incorporated by electrospinning process.

*Table 3 Mode I fracture energy for each material type.*

| Material type | Mode I fracture energy |
|---|---|
| Reference | 0.33 |
| BMI modified (SEP) | 0.75 |
| BMI-GNP modified (SEP) | 0.89 |
| BMI modified (MEP) | 0.75 |
| BMI-GNP modified (MEP) | 0.89 |

*Table 4 Hashin's damage initiation properties for all modified material types.*

|  | BMI modified (SEP) | BMI-GNP modified (SEP) | BMI modified (MEP) | BMI-GNP modified (MEP) |
|---|---|---|---|---|
| **Longitudinal tensile strength $X_t$ (MPa)** | 2750 | 2750 | 2750 | 2750 |
| **Longitudinal compressive strength $X_c$ (MPa)** | 1950 | 1950 | 1950 | 1950 |
| **Transverse tensile strength $Y_t$ (MPa)** | 30 | 200 | 30 | 300 |

| Transverse compressive strength $Y_c$ (MPa) | 100 | 269 | 100 | 370 |
| Longitudinal shear strengths $S_{12} = S_{13}$ (MPa) | 40 | 40 | 40 | 40 |
| Transverse shear strength $S_{23}$ (MPa) | 90 | 90 | 90 | 90 |

*Table 5 Hashin's damage evolution properties for all modified material types.*

|  | BMI modified (SEP) | BMI-GNP modified (SEP) | BMI modified (MEP) | BMI-GNP modified (MEP) |
| --- | --- | --- | --- | --- |
| Longitudinal tensile fracture energy (*mJ*) | 400 | 400 | 400 | 400 |
| Longitudinal compressive fracture energy (*mJ*) | 74 | 74 | 74 | 74 |
| Transverse tensile fracture energy (*mJ*) | 1 | 1 | 1 | 1 |
| Transverse compressive fracture energy (*mJ*) | 18 | 18 | 18 | 18 |

Table 6 Cohesive contact damage initiation properties for all modified material types.

|  | BMI modified (SEP) | BMI-GNP modified (SEP) | BMI modified (MEP) | BMI-GNP modified (MEP) |
| --- | --- | --- | --- | --- |
| Normal strength (*MPa*) | REF: 300 SHA: 200 | REF: 300 SHA: 200 | REF: 300 SHA: 200 | REF: 300 SHA: 200 |
| Shear-1 strength (*MPa*) | REF: 150 SHA: 150 | REF: 150 SHA: 150 | REF: 150 SHA: 50 | REF: 150 SHA: 30 |

| Shear-2 strength ($MPa$) | REF: 150 SHA: 150 | REF: 150 SHA: 150 | REF: 150 SHA: 50 | REF: 150 SHA: 30 |

Overall, the modeling framework established in this study combines high-resolution finite element analysis with realistic material and interface behavior, enabling robust predictions of failure in self-healing open-hole CFRP laminates under tensile loading. The following sections present and analyze the simulation results in the context of the observed experimental phenomena.

*Table 7 Cohesive contact damage evolution properties for all modified material types.*

|  | BMI modified (SEP) | BMI-GNP modified (SEP) | BMI modified (MEP) | BMI-GNP modified (MEP) |
|---|---|---|---|---|
| **Normal fracture energy ($mJ$)** | REF: 0.33 SHA: 0.75 | REF: 0.33 SHA: 0.89 | REF: 0.33 SHA: 0.75 | REF: 0.33 SHA: 0.89 |
| **1st Shear fracture energy ($mJ$)** | REF: 1 SHA: 1 | REF: 1 SHA: 1 | REF: 1 SHA: 1 | REF: 1 SHA: 1 |
| **2nd Shear fracture energy ($mJ$)** | REF: 1 SHA: 1 | REF: 1 SHA: 1 | REF: 1 SHA: 1 | REF: 1 SHA: 1 |
| **BK criterion exponent** | REF: 1.2 SHA: 1.2 | REF: 1.2 SHA: 1.2 | REF: 1.2 SHA: 1.2 | REF: 1.2 SHA: 1.2 |

## 3 Results And Discussion

### Load-Displacement Behavior

The numerical simulations produced load-displacement responses that closely mirrored the experimentally observed trends, particularly during the initial elastic phase and the early stages of damage evolution, Figure 7. In all configurations, the response was initially linear, followed by a deviation corresponding to matrix cracking and delamination initiation. Notably, the SEP-modified specimens exhibited higher peak loads and greater post-yield deformation capacity compared to their MEP and unmodified counterparts, indicating improved toughness and damage tolerance due to the uniformly distributed self-healing agent.

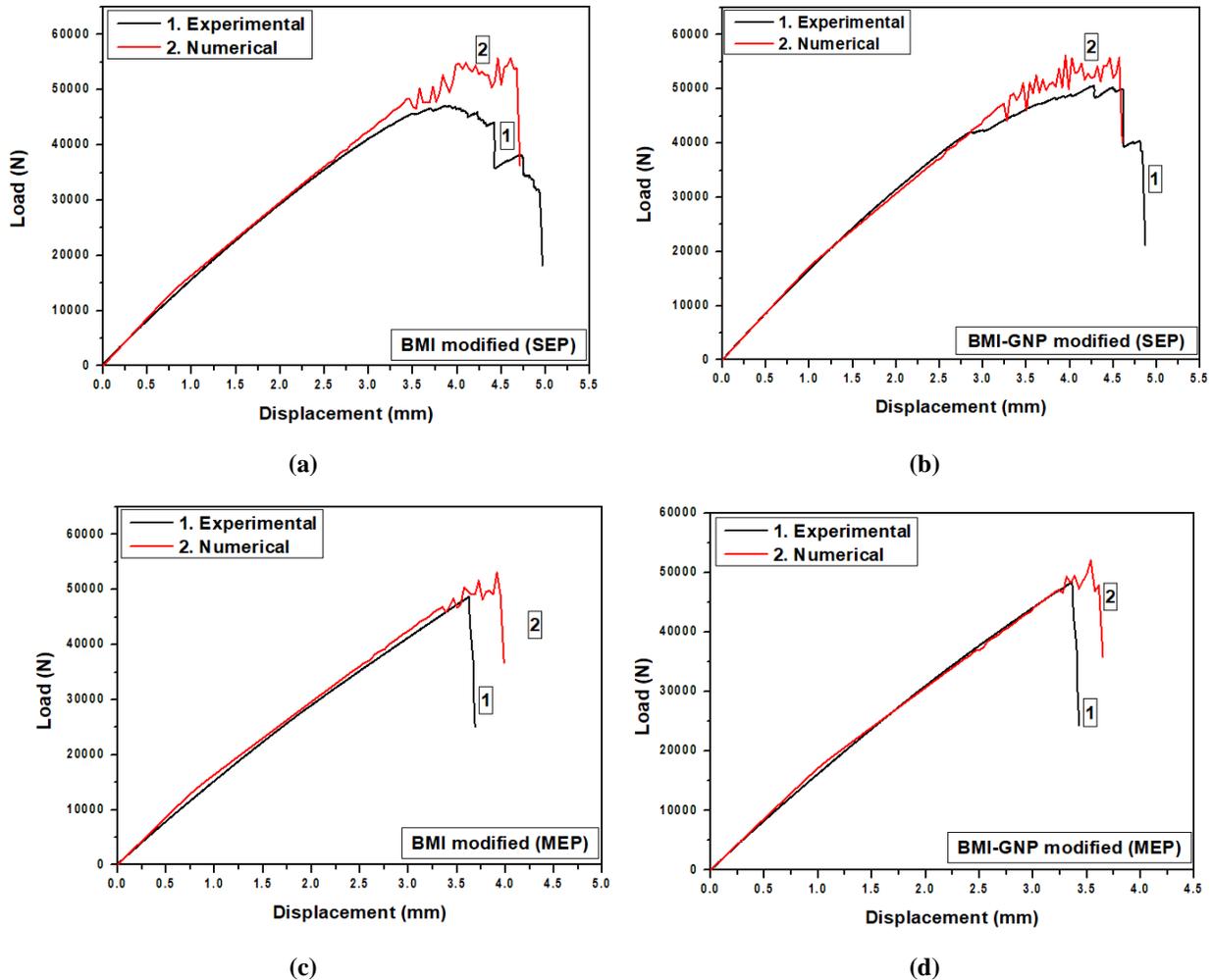

*Figure 7: Numerical and representative experimental load-displacement curves (a) BMI modified (SEP), (b) BMI-GNP modified (SEP), (c) BMI modified (MEP) and (d) BMI-GNP modified (MEP).*

BMI Modified (SEP)

The numerical simulation of the BMI-modified laminate using solution electrospinning (SEP) showed strong alignment with the experimental load-displacement response, particularly in the linear region up to 2.6 mm displacement, Figure 7 (a). Beyond this point, the experimental curve deviated gradually due to damage initiation, displaying two characteristic load drops: the first around 4.4 mm likely due to delamination initiation, and the second near 5 mm corresponding to complete material failure. In contrast, as seen in Figure 8, the numerical model predicted a later deviation (around 3.5 mm) and exhibited higher load-bearing capacity with a more abrupt failure at approximately 4.7 mm. Although the model did not replicate the dual load drops, it captured the overall response and damage progression effectively.

Damage mapping from the simulation revealed matrix tensile damage initiating symmetrically around the hole and progressing toward the tabs after 3.4 mm displacement. Fiber tension damage remained confined to the hole vicinity, while compressive fiber and matrix damage appeared near the tabs and progressed post-fracture. Inter-ply delaminations initiated near the tabs and expanded toward the center as load increased. Visual comparison with tested specimens confirmed the presence of matrix cracks, fiber pull-out, and minor delaminations in the tabs, closely matching the numerical predictions.

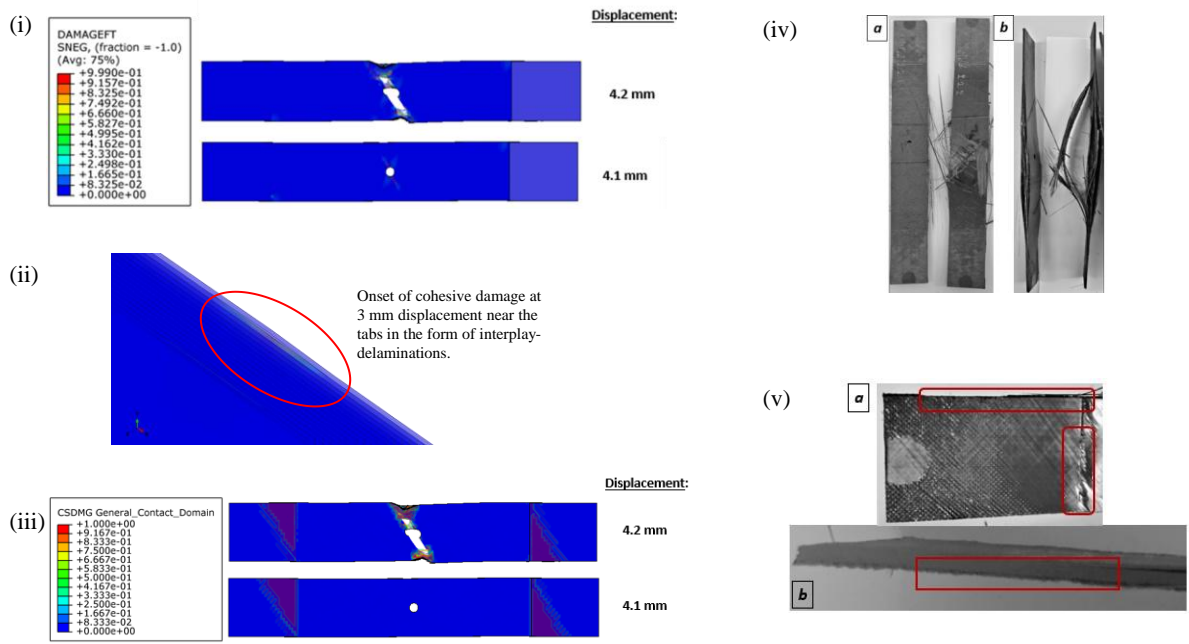

*Figure 8 BMI modified (SEP)- (i) Fiber tension damage at 4.1 mm and 4.2 mm displacement, (ii) Onset of cohesive damage at 3 mm displacement near the tabs in the form of interplay-delaminations, (iii) Cohesive contact damage at 4.1 mm and 4.2 mm displacement, (iv) Damage in BMI modified by SEP specimens after quasi-static tensile testing: (a) front view (b) side view and (v) Close-up of delaminations and damage in the end-tab region of the BMI modified by SEP specimens after quasi-static tensile testing: (a) front view, (b) side view*

BMI-GNP Modified (SEP)

For laminates modified with graphene nanoplatelet (GNP)-enhanced BMI via SEP, the numerical and experimental load-displacement curves remained closely aligned through the elastic region, Figure 7 (b). The model accurately captured the onset of instability and a primary load drop around 4.1 mm displacement, though it did not reproduce the secondary experimental load drop. From Figure 9 it can be observed that damage again initiated symmetrically around the hole and expanded outward. Fiber tension and matrix compression damage localized near the hole, while cohesive failures began in the tabs and spread centrally. Experimental observations also showed matrix cracking, fiber pull-out, and delaminations concentrated around the hole, validating the numerical results.

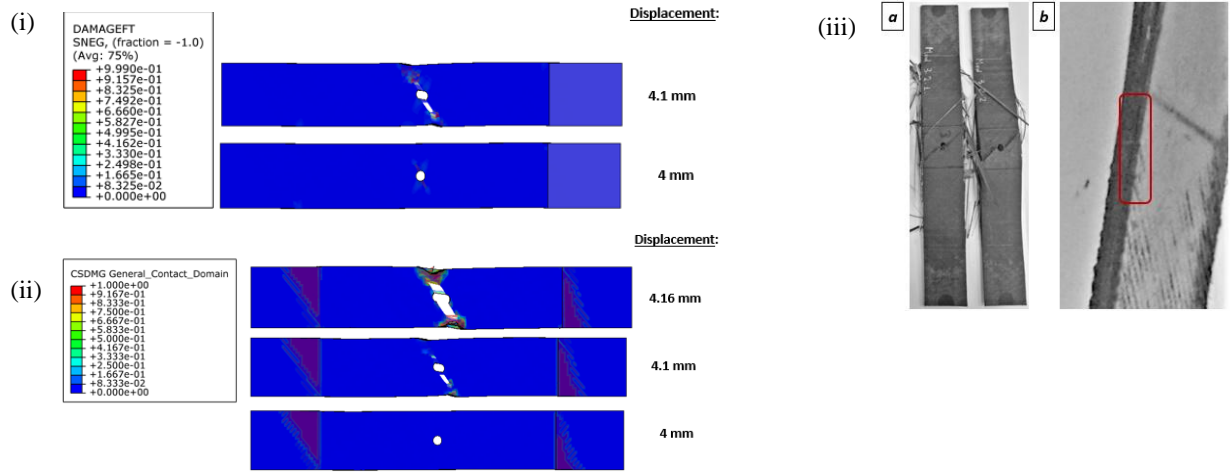

*Figure 9 BMI-GNP modified (SEP)- (i)Fiber tension damage at 4.1 mm and 4 mm displacement, (ii) Cohesive contact damage at various displacements and (iii) Damage in BMI-GNP modified by SEP specimens after quasi-static tensile tests: (a) front view (b) close-up at tab region*

BMI Modified (MEP)

Simulations of BMI-modified laminates produced via melt electrospinning (MEP) showed reasonable agreement with experimental results, Figure 7 (c). The numerical model demonstrated a higher peak load and delayed failure compared to experiments, with instabilities beginning around 3.4 mm displacement. Fiber tension and compressive failures occurred near the center, and delaminations were concentrated around the hole, Figure 10. Unlike SEP-based models, damage in the tabs was minimal, both numerically and experimentally. The central concentration of damage formed an hourglass pattern consistent with specimen observations.

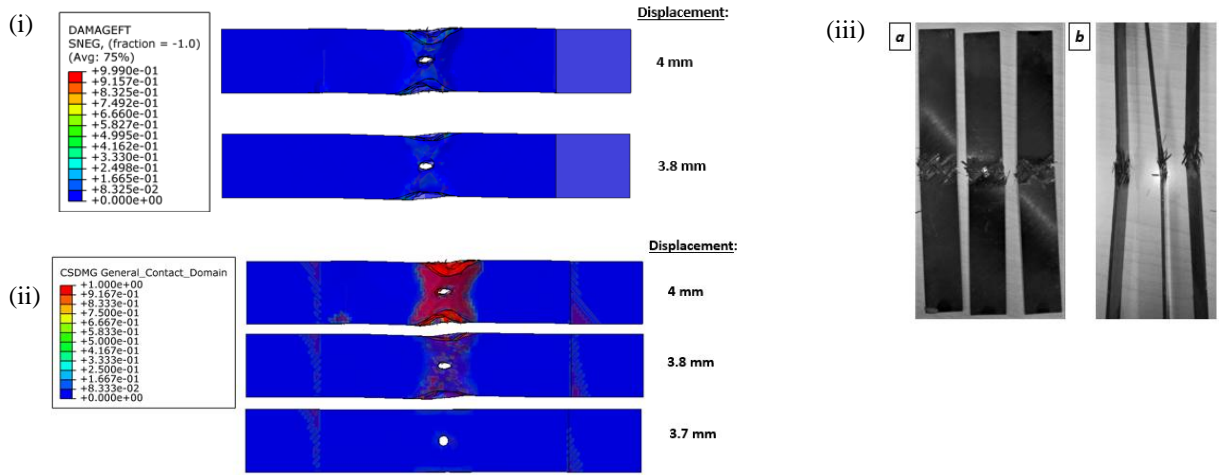

*Figure 10 BMI modified (MEP)- (i) Fiber tension damage at 4 mm and 3.8 mm displacement, (ii) Cohesive contact damage at various displacements and (iii) Damage in BMI modified by MEP specimens* after quasi-static tensile tests*: (a) front view, (b) side view.*

BMI-GNP Modified (MEP)

The numerical results for GNP-modified BMI laminates fabricated by MEP showed an initially linear response closely matching experimental data, Figure 7 (d). Both curves peaked around the same load level, followed by a load drop due to delamination around the hole. Fiber and matrix damage remained localized, and cohesive failures began in the tabs but shifted centrally as loading progressed, Figure 11. Post-test images of the specimens revealed clear delamination around the hole with minimal tab damage, aligning well with simulation outputs.

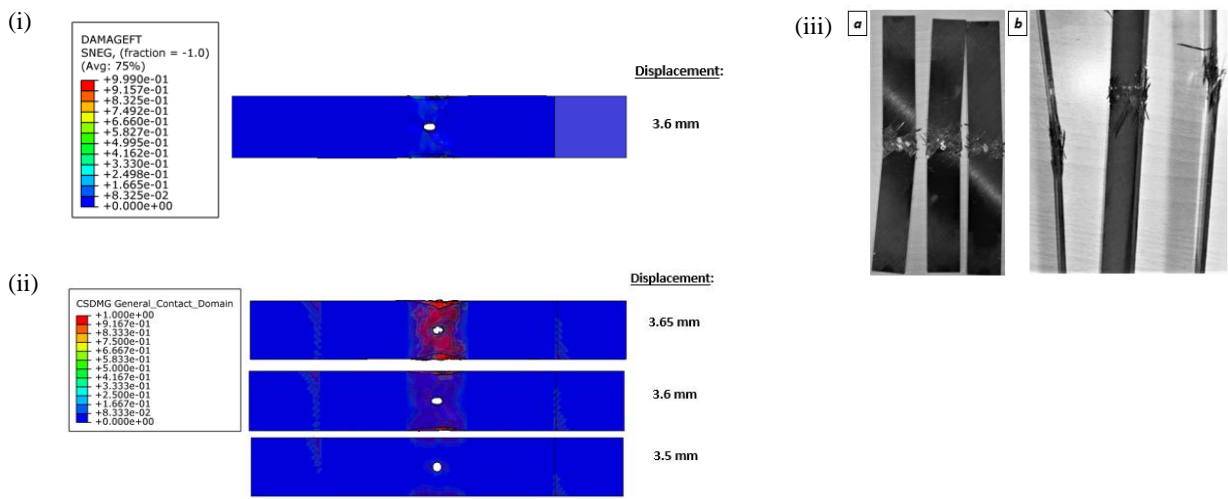

*Figure 11* BMI-GNP modified (MEP)- (i) Fiber tensile damage at 3.6 mm displacement, (ii) Cohesive contact damage at various displacements and (iii) *Damage in BMI-GNP modified by MEP specimens* after quasi-static tensile tests*: (a) front view, (b) side view.*

Comparative Summary

Table 8 summarizes the displacements corresponding to the onset of cohesive, matrix, and fiber damage. It shows that the onset of damage is consistent within the same SHA type, with SEP-modified materials generally initiating failure along the width of the hole, while MEP-modified ones fail primarily through delamination. Table 9 compares experimental and numerical maximum loads. Deviations ranged from 9.3% to 12.6%, with SEP-GNP models exhibiting the highest overestimation. While overall agreement was acceptable, the results indicate opportunities for refining damage progression algorithms and improving the accuracy of cohesive behavior representation.

*Table 8 Onset of cohesive, matrix tension and fiber tension damage in terms of displacement and load for all material types.*

| Material type | Onset of cohesive damage (mm) | | Onset of matrix tension damage (mm) | | Onset of fiber tension damage (mm) | |
|---|---|---|---|---|---|---|
| | [mm] | [kN] | [mm] | [kN] | [mm] | [kN] |
| BMI modified (SEP) | 3 | 42.5 | 2.3 | 33.6 | 4.2 | 54.2 |
| BMI-GNP modified (SEP) | 2.9 | 42.7 | 1.9 | 29.4 | 4.1 | 53.4 |
| BMI modified (MEP) | 3 | 42.5 | 2.3 | 33.6 | 3.8 | 49.6 |
| BMI-GNP modified (MEP) | 2.9 | 42.6 | 1.9 | 29.4 | 3.6 | 47.6 |

Based on the above, the MEP-modified specimens demonstrated a more localized damage response, with cohesive failures occurring primarily in the regions reinforced by the electrospun interleaves. This led to a more abrupt load drop after peak, as the reinforcement was not present throughout the laminate. The reference (unmodified) samples consistently showed earlier onset of stiffness degradation and lower maximum load capacity, confirming the mechanical benefits of the integrated SHAs.

Table 9 Deviation of experimental and analytically computed maximum loads for all modified OH CFRP models.

| Material type | Experimental mean maximum load | Analytical maximum load | Deviation |
|---|---|---|---|
| | [N] | [N] | [%] |
| BMI modified (SEP) | 50227 | 55742 | 11.0 |
| BMI-GNP modified (SEP) | 49527 | 55768 | 12.6 |
| BMI modified (MEP) | 48243 | 53094 | 10.1 |
| BMI-GNP modified (MEP) | 47665 | 52100 | 9.3 |

Damage Initiation and Propagation Numerical predictions accurately identified the locations of initial matrix cracking and interlaminar delamination, particularly near the open hole and tab regions. Matrix damage in tension initiated symmetrically around the notch, gradually expanding toward the specimen edges as loading increased. Fiber breakage was confined mostly to the regions adjacent to the hole, consistent with high stress concentrations.

Cohesive contact failures—interpreted as delaminations—began at the inner plies in the tab areas and subsequently spread toward the center of the laminate. This progression was more pronounced in SEP-modified models, where increased interface strength delayed delamination onset but allowed wider delamination zones once initiated. MEP-modified models demonstrated more controlled and localized damage growth, confirming the effectiveness of targeted reinforcement.

Visual inspection of simulated failure patterns closely resembled those observed in tested specimens. The SEP-modified configurations exhibited extensive matrix cracking and delamination throughout the gage length, aligning with photographic evidence of fractured surfaces. MEP-modified specimens showed damage confined to regions surrounding the hole and had minimal tab-region failures, again matching experimental findings.

The cohesive damage variable evolution provided further insight into interfacial degradation. In all simulations, damage variables increased rapidly after the peak load, especially in zones of material heterogeneity. The use of differential cohesive properties for modified and unmodified areas proved critical in capturing this behavior, highlighting the importance of accurate spatial representation of SHA distribution in the numerical model.

Despite the strong qualitative agreement, certain discrepancies between numerical and experimental results were noted. Specifically, the simulations tended to overpredict the peak load, likely due to idealized material parameters and uniform element behavior. Additionally, the abruptness of final failure was more pronounced in the models, a known limitation of explicit simulations lacking damage stabilization.

To further improve fidelity, future models could implement radial meshing near the notch to better resolve stress gradients and damage trajectories. Incorporating rate-dependent damage models and implementing intraply matrix cracking explicitly could also enhance accuracy. The inclusion of fatigue-based delamination growth models is recommended for long-term performance prediction.

In summary, the developed finite element models effectively captured the primary failure mechanisms observed experimentally in open-hole CFRP laminates with and without self-healing modifications. The results affirm the mechanical advantages of electrospun interleaves and provide a strong foundation for predictive modeling of advanced multifunctional composites.

## 4  Conclusions

This study successfully developed and validated finite element models for predicting damage evolution in open-hole carbon fiber/epoxy composites modified with electrospun self-healing agents. By integrating Hashin's failure criteria for intralaminar damage and surface-based cohesive contact formulations for interlaminar delamination, the simulations captured the complex progression of matrix cracking, fiber breakage, and delamination observed during quasi-static tensile loading.

The inclusion of Diels-Alder based self-healing interleaves significantly enhanced the mechanical performance of the laminates. Solution electrospinning (SEP) provided distributed reinforcement across the laminate thickness, resulting in higher load capacity and more extensive damage tolerance. In contrast, melt electrospinning (MEP) enabled more localized toughening, which helped control the damage evolution around stress concentration zones without excessively increasing weight or altering the laminate's global stiffness.

The comparative analysis between simulations and experimental results demonstrated good correlation in terms of load-displacement behavior and damage morphology. The model accurately identified failure initiation sites and was able to reproduce key failure features, including inter-ply delamination and matrix cracking patterns.

However, limitations in the current modeling approach were acknowledged. The simulations tended to overestimate the peak load, possibly due to idealized material homogeneity and simplified cohesive property definitions. Furthermore, the absence of mechanisms such as longitudinal matrix splitting and fatigue crack propagation constrained the model's ability to fully capture post-peak and long-term behavior.

Overall, this work demonstrates the feasibility and value of integrating numerical modeling with self-healing material concepts for improving the reliability and lifespan of aerospace-grade composites. The methodology presented provides a foundation for advanced simulation frameworks capable of supporting design and certification efforts for next-generation smart composite structures.